\documentclass[conference]{IEEEtran}

\usepackage{amsmath}
\usepackage{amsthm}
\usepackage{amssymb}
\usepackage{graphicx}
\usepackage{epsfig}
\usepackage{algorithmicx}  
\usepackage{algpseudocode}  
\usepackage{algorithm}
\usepackage{multirow}
\usepackage{subcaption}
\usepackage{setspace}

\begin{document}

\title{Overlap-Summation-Based Pulse Shaping Transceiver for Affine Frequency Division Multiplexing}

\author{\IEEEauthorblockN{Haojian Zhang\IEEEauthorrefmark{1}, Jiayan Yang\IEEEauthorrefmark{1}, Tingting Zhang\IEEEauthorrefmark{1}, Xu Zhu\IEEEauthorrefmark{1}, Qinyu Zhang\IEEEauthorrefmark{1}
\IEEEauthorblockA{\IEEEauthorrefmark{1} School of Information Science and Technology, Harbin Institute of Technology (Shenzhen), Shenzhen 518055, China}
Email: zhangtt@hit.edu.cn}}

\maketitle
\IEEEpeerreviewmaketitle

\begin{table*}[!t]
	\centering
	\tiny
	\begin{equation}
		y\left[m\right]=\sum_{m^{\prime}=0}^{M-1}x\left[m^{\prime}\right]\left(\overbrace{\sum_{p=0}^{P-1}\underbrace{\sum_{\ell^{\prime}=\lfloor\ell_{p}\rfloor-L_{\text{T}}/2}^{\lceil\ell_{p}\rceil+L_{\text{T}}/2}h_{p}e^{\jmath 2\pi\left(c_{1}{\ell^{\prime}}^{2}+c_{2}\left({m^{\prime}}^{2}-m^{2}\right)-\frac{m^{\prime}\ell^{\prime}}{M}\right)}g_{\text{T}}\left(\Delta_{\text{T}}\left(\ell^{\prime}-\ell_{p}\right)\right)g_{\text{W}}\left(\Delta_{\text{F}}\left(m-\left(m^{\prime}-\left(2Mc_{1}\ell^{\prime}-k_{p}\right)\right)\right)\right)}_{g_{\text{X}}\left[m,m^{\prime},p\right]}}^{\left[\mathbf{H}\right]_{m,m^{\prime}}}\right)+w\left[m\right].\label{eq:inout}
	\end{equation}
\end{table*}
\begin{figure*}[!t]
	\centering
	\includegraphics[scale=0.14]{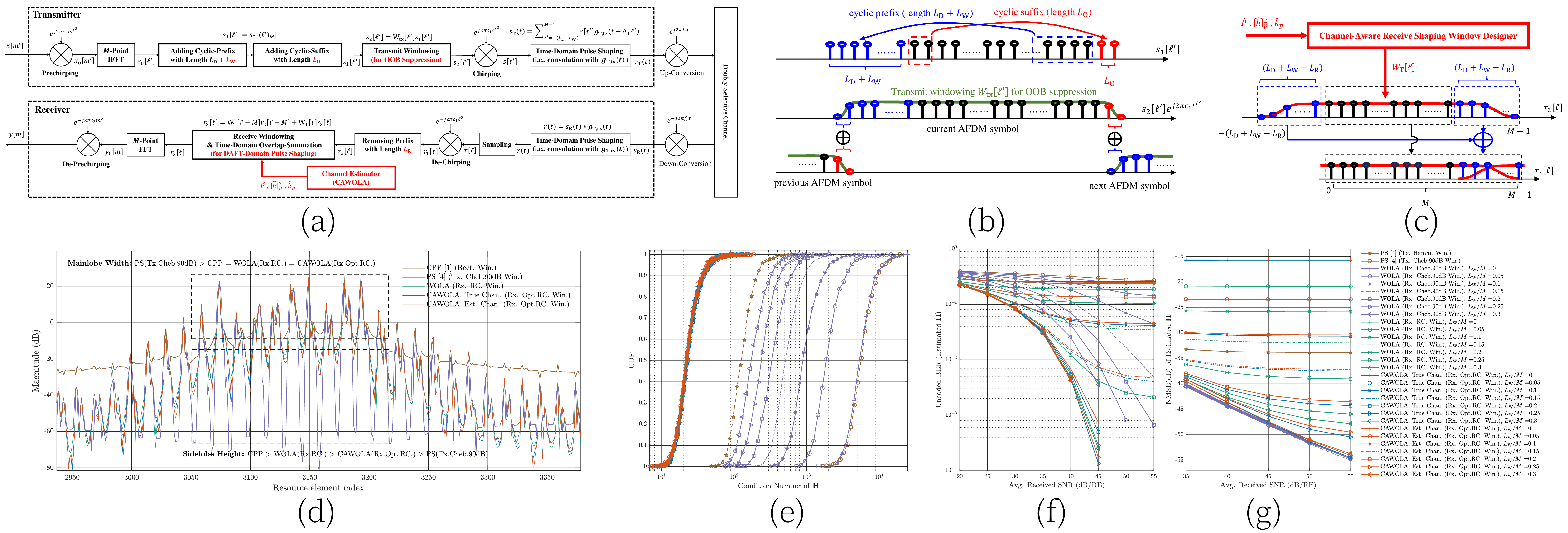}
	\caption{Models and results of the proposed CAWOLA-AFDM transceiver, (a) overall block diagram; (b) sketch of the transmit WOLA operation; (c) sketch of the receive CAWOLA operation; (d) one random DAFT-domain profile of the received embedded pilot; (e) channel condition number; (f) uncoded BER; (g) channel-estimation NMSE. The enlarged figures are provided in the supplementary file.}
	\label{fig:psRxTx}
\end{figure*}
\noindent Affine frequency division multiplexing (AFDM) has recently emerged as a promising waveform for doubly-selective channels~\cite{ref_afdm,ref_chirpSignal,ref_afdm4}. In \cite{ref_pulse}, a direct-windowing-based pulse shaping transceiver (PS-AFDM) was proposed to suppress the Doppler sidelobes, thus improving the accuracy of channel estimation. We observe that, when the path delays and Doppler shifts are randomly distributed, the legacy PS-AFDM scheme in \cite{ref_pulse} significantly increases the condition number of the effective channel matrix. The resulting ill-conditioning degrades the numerical stability of channel equalization in noise and consequently increases the BER. To address this issue, this letter applies the existing weighted overlap-summation (WOLA) transceiver to AFDM and proposes a novel \emph{channel-aware} (CA) receive shaping window design, which simultaneously achieves accurate channel estimation and robust equalization performance, at the cost of additional prefix overhead and receive-window calculation. The resulting scheme is termed CAWOLA-AFDM. Compared with the legacy WOLA scheme in \cite{ref_wola}, which employs a fixed receive window, the proposed CAWOLA design exploits the channel estimates of slowly varying power gains and Doppler shifts to design a channel-tailored receive shaping window in closed form, thereby further enhancing channel-estimation accuracy while maintaining the channel condition number when a Nyquist prototype window is adopted. The proposed CAWOLA receive-window design aims to produce an effective AFDM channel with more compact support in the DAFT domain. The source code for the simulations is provided at https://github.com/SANIS-HITSZ/Waveform\_AFDM.

\section{WOLA-based AFDM transceiver}
The architecture and input-output (I/O) relationship of the generalized WOLA-based transceiver for AFDM, both of which are retained in the proposed CAWOLA-AFDM, are first introduced in this section. The key novelty of the proposed CAWOLA lies in exploiting the AFDM channel information to design a channel-tailored receive shaping window $W_{\text{T}}\left[\ell\right]$. Fig.~\ref{fig:psRxTx}(a) shows the block diagram of the WOLA-based AFDM transceiver. In particular, $L_{\text{O}}$ and $W_{\text{tx}}\left[\ell^{\prime}\right]$ denote the suffix length and transmit window, respectively, for out-of-band (OOB) emission suppression. $L_{\text{W}}$ and $W_{\text{T}}\left[\ell\right]$ denote the extended prefix length and receive window for DAFT-domain pulse shaping, respectively. $\widehat{P}$, $\widehat{\left|h\right|}^{2}_{p}$ and $\widehat{k}_{p}$ denote the estimated path number, power gain, and off-grid Doppler shift, respectively. The illustrations of transmit WOLA and receive CAWOLA operations in the proposed scheme are shown in Fig.~\ref{fig:psRxTx}(b) and (c), respectively. Eqn.~(\ref{eq:inout}) shows the I/O relationship of the WOLA-based transceiver, while the detailed derivations are given in Appendix~A. The purpose of introducing the WOLA-based transceiver into AFDM systems is to adjust the sidelobes of DAFT-domain pulse $g_{\text{X}}\left[m,m^{\prime},p\right]$ while maintaining the Doppler resolution at $\Delta_{\text{F}}$. We observe that the RC-window-based WOLA transceiver maintains a lower channel condition number than PS-AFDM \cite{ref_pulse}. The proposed CAWOLA design further improves channel-estimation accuracy. Note that $\ell_{p}$ and $k_{p}$ are both fractional numbers in Eqn.~(\ref{eq:inout}). $g_{\text{T}}\left(t\right)$ denotes the overall time-domain shaping pulse. $g_{\text{W}}\left(f\right)$ denotes the discrete-time Fourier transform of the shaping window $W_{\text{T}}\left[\ell\right]$ at a sampling rate of $M\Delta_{\text{F}}$, given by
\begin{equation}
	g_{\text{W}}\left(f\right)=\frac{1}{M}\sum\nolimits_{\ell=-\left(L_{\text{D}}+L_{\text{W}}-L_{\text{R}}\right)}^{M-1}W_{\text{T}}\left[\ell\right]e^{-\jmath\left(\frac{2\pi f}{M\Delta_{\text{F}}}\right)\ell}.\label{eq:gw}
\end{equation}
The definitions of symbols are provided in Appendix~A.

\section{The proposed CAWOLA shaping design} 
The motivation of the proposed CAWOLA shaping design is explained in Appendix~B.1. Given a real-valued prototype receive window $W_{\text{T}}^{\mathfrak{R}}\left[\ell\right]$ (i.e., the receive window employed in the legacy WOLA), the proposed CAWOLA receive shaping window is expressed as
\begin{equation}
	\footnotesize
  	W_{\text{T}}\left[\ell\right]=\left\{
		\begin{aligned}
			&a_{0}W_{\text{T}}^{\mathfrak{R}}\left[\ell\right]+\sum_{n=0}^{N-1}a_{n+1}\omega_{n}\left[\ell+L\right]~,~-L\leq\ell<0,\\
			&a_{0}W_{\text{T}}^{\mathfrak{R}}\left[\ell\right]~,~0\leq\ell<M-L,\\
			&a_{0}W_{\text{T}}^{\mathfrak{R}}\left[\ell\right]+\sum_{n=0}^{N-1}a_{n+N+1}\omega_{n}\left[\ell+L-M\right]~,~M-L\leq\ell<M,
		\end{aligned}
  	\right.\label{eq:win}
\end{equation}
where
\begin{equation}
	\omega_{n}\left[\ell\right]=\left(\frac{\ell}{L-1}\right)\left(1-\frac{\ell}{L-1}\right)\cos\left(n\pi\frac{\ell}{L-1}\right).
\end{equation}
The proposed CAWOLA employs a set of cosine-like basis functions over its leading and trailing $L$ samples to reshape the Doppler response, which is motivated by the low-order energy-compaction property of the discrete cosine transform. Moreover, the envelope $\left(\ell/\left(L-1\right)\right)\left(1-\ell/\left(L-1\right)\right)$ is applied to each basis function to force the linear combination to vanish at both endpoints, thereby preserving the continuity of $W_{\text{T}}\left[\ell\right]$.
\par To determine the coefficient $a$, the power profile of the $p$-th path associated with the delay grid $\ell^{\prime}$ is defined as
\begin{equation}
  G_{p,\ell^{\prime}}\left[q\right]=\left|h_{p}g_{\text{T}}\left(\Delta_T(\ell^{\prime}-\ell_{p})\right)\right|^{2}\left|g_{\text{W}}\left(\Delta_{\text{F}}\left(q-k_{p}\right)\right)\right|^{2},
  \label{eq:G}
\end{equation}
according to the I/O relationship in Eqn.~(\ref{eq:inout}), where $-M/2\leq q<M/2$. The term $\left(m-\left(m^{\prime}-\left(2Mc_{1}\ell^{\prime}-k_{p}\right)\right)\right)$ in the I/O relationship indicates that a spacing of $2Mc_{1}$ bins is reserved between adjacent delay-grid indices $\ell^{\prime}$ in the DAFT-domain, thereby separating their corresponding Doppler spreads. Accordingly, this letter proposes a method for determining the window coefficient $a$ by maximizing the path-weighted energy ratio of the power profile confined to the reserved band of $-\left(2Mc_{1}-1\right)/2\leq q\leq\left(2Mc_{1}-1\right)/2$, i.e.,
\begin{equation}
	\footnotesize
	\begin{bmatrix}
		a_{0}^{\star}\\
		\vdots\\
		a_{2N}^{\star}
	\end{bmatrix}=\mathop{\arg\max}_{a_{0},\cdots,a_{2N}}\frac{\sum\limits_{p=0}^{P-1}\sum\limits_{\ell^{\prime}=\lfloor\ell_{p}\rfloor-L_{\text{T}}/2}^{\lceil\ell_{p}\rceil+L_{\text{T}}/2}\sum\limits_{q=-\left(2Mc_{1}-1\right)/2}^{\left(2Mc_{1}-1\right)/2}G_{p,\ell^{\prime}}\left[q\right]}{\sum\limits_{p=0}^{P-1}\sum\limits_{\ell^{\prime}=\lfloor\ell_{p}\rfloor-L_{\text{T}}/2}^{\lceil\ell_{p}\rceil+L_{\text{T}}/2}\sum\limits_{q=-M/2}^{M/2-1}G_{p,\ell^{\prime}}\left[q\right]}.\label{eq:opt0}
\end{equation}
The above problem can be reformulated in vector form as
\begin{equation}
	\mathbf{a}^{\star}=\mathop{\arg\max}_{\mathbf{a}}\left.\left(\mathbf{a}^{\mathrm{H}}\mathbf{R}_{\text{inn}}\mathbf{a}\right)\middle/\left(\mathbf{a}^{\mathrm{H}}\mathbf{R}_{\text{all}}\mathbf{a}\right)\right..\label{eq:opt}
\end{equation}
The optimal $\mathbf{a}^{\star}$ is given in closed form by
\begin{equation}
	\mathbf{a}^{\star}=\mathop{\text{generalized eigenvector}}_{\lambda_{\max}}\left\{\mathbf{R}_{\text{inn}},\mathbf{R}_{\text{all}}+\frac{\epsilon\operatorname{Tr}\left\{\mathbf{R}_{\text{all}}\right\}}{2N+1}\mathbf{I}\right\},\label{eq:solve}
\end{equation}
where $\epsilon$ is a regularization factor introduced to improve numerical stability in practice. The CAWOLA receive shaping window $W_{\text{T}}\left[\ell\right]$ can be constructed from Eqn.~(\ref{eq:win}) using the normalized coefficient vector $\mathbf{a}^{\star}/a_{0}^{\star}$. Detailed derivations, computational steps, and complexity analyses of the proposed CAWOLA design are summarized in Appendix~B.

\section{Numerical results}
The proposed CAWOLA is evaluated in terms of the condition number of $\mathbf{H}$, the channel-estimation NMSE using the method in \cite{ref_afdm}, and the uncoded BER using an LMMSE equalizer. We consider a doubly-selective channel with the parameters shown in Appendix~C. As shown in Fig.~\ref{fig:psRxTx}(d)-(g), for $M_{\text{schd}}$=600, $P$=10, $L_{\max}$=10 and $K_{\max}$=3, employing a Chebyshev window yields lower DAFT-domain sidelobes, thereby enabling high-accuracy channel estimation. However, it also causes pronounced mainlobe widening, which increases the interference between adjacent modulated symbols. Consequently, the correlation between adjacent columns of $\mathbf{H}$ is observed to increase, which raises the channel condition number and makes channel equalization more sensitive to noise, ultimately degrading the BER performance. Compared with WOLA, the RC-window-based CAWOLA scheme further suppresses the sidelobes while preserving the mainlobe width. At the same prefix overhead as WOLA, CAWOLA therefore enables high-accuracy channel estimation without increasing the channel condition number, ultimately yielding a low BER. More simulation results are provided in Appendix~C.

\section{Conclusion}
This letter has proposed a CAWOLA-AFDM transceiver that achieves accurate channel estimation while maintaining a low condition number of the channel matrix when a Nyquist prototype window is adopted. In future work, the influence of pulse shaping on the condition number of the channel matrix requires further mathematical analysis and rigorous proof.

\section*{Acknowledgment}
The authors would like to thank Dr. Qu Luo for his careful reading and valuable suggestions.


\begin{thebibliography}{99}
	
	\bibitem{ref_afdm}
	Bemani A, Ksairi N, Kountouris M. Affine frequency division multiplexing for next generation wireless communications. IEEE Trans Wireless Commun, 2023, 22(11): 8214--8229

	\bibitem{ref_afdm4}
	Ranasinghe K, Ge Y, Abreu G. Joint channel estimation and data detection for AFDM receivers with oversampling. ICNC 2025, Honolulu, HI, USA, 2025

	\bibitem{ref_chirpSignal}
	Sui Z, Luo Q, Liu Z, et al. Multi-functional chirp signalling for next-generation multi-carrier wireless networks: communications, sensing and ISAC perspectives. arXiv preprint, arXiv:2508.06022

	\bibitem{ref_pulse}
	Yin H R, Tang Y Q, Li S Y, et al. Evaluation and design criterion for pulse-shaped AFDM. In: IEEE GLOBECOM 2024, Cape Town, South Africa, 2024

	\bibitem{ref_wola}
	Zayani R, Medjahdi Y, Shaiek H, et al. WOLA-OFDM: a potential candidate for asynchronous 5G. In: IEEE GLOBECOM 2016, Washington DC, USA, 2016

\end{thebibliography}
\end{document}